\documentclass[prb,twocolumn,aps]{revtex4}
\usepackage{graphicx}
\usepackage{bm}
\usepackage{epsfig}
\usepackage{psfrag}
\begin{document}


\title{Electromagnetomotive force fields in noninertial reference frames and\\
accelerated superconducting quantum interferometers}
\author{Uwe R. Fischer, Christoph H\"aussler, J\"org Oppenl\"ander, and Nils Schopohl} 
\affiliation{Eberhard-Karls-Universit\"at T\"ubingen, Institut f\"ur Theoretische Physik \\
Auf der Morgenstelle 14, D-72076 T\"ubingen, Germany}

\date{\today}


\begin{abstract}
We discuss the prospects of detecting with high precision the force fields
related to noninertiality in superconducting circuits. 
Special emphasis is laid on the perfectly conducting and perfect diamagnetism
analogues of the Tolman-Stewart respectively Barnett effects. 
The influence of acceleration and rotation
on the electrodynamics of superconducting interferometers 
is explicitly described.
In particular, we show how motion induced changes of the oscillation frequency
of the local Josephson oscillators in superconducting quantum interference filters
can be used for precision measurements of acceleration in free space.
\end{abstract}
\vspace*{0.6cm}
\maketitle
\section{Introduction}
The detection of electromagnetic fields induced by acceleration in magnetizable materials
and metals has a history reaching back more than one century \cite{JCMaxwell}. 
Two famous experiments stand out in this respect. In 1915, Barnett 
measured the magnetic field of a magnetizable material induced
by rotating it \cite{barnett}. One year later, Tolman and Stewart measured 
the electromotive force if metals are (linearly) accelerated \cite{tolmanstewart1}. 
They thereafter concluded on the effective mass of the current carrier 
\cite{tolmanstewart2}, which turned out be 
somewhat different from that of the electron {\it in vacuo}. 
A more modern experiment  with increased precision
in rotationally accelerated conductors 
has been carried out \cite{MoorheadOpat}, 
where it was found that the mass is the electron 
mass in vacuum to within about one percent \cite{REVMODPHYS}.

Superconductors and, more generally, macroscopically coherent quantum 
systems, are distinguished by the fact that the mass of the current
carriers has been measured, within currently achievable precision of a 
few ppm to be {\em exactly} twice the bare mass $m_e$ of the electrons {\it in vacuo}.
The mass is subject to very small (relativistic) corrections only, so small 
as to currently elude precise experimental determination.
The measurement of the bare mass proceeds via the  
magnetomechanical effect in superconductors,
the London moment \cite{London1,zimmerman,londoneqliu}. 
The London magnetic field induced by rotation (cf. equation (\ref{London}) below) 
is proportional to the rotation velocity and 
to the ratio of twice the mass $m$ of the superconducting 
current carriers divided by their charge $q$. The best measurement 
to date \cite{tatecabrera} yielded for the current carrying Cooper pairs 
a mass $m/2m_e=1.000084(21)$. 

The London moment is a universal magnetomechanical 
property of rotating superconductors, independent of specific material properties, and 
verified not only in conventional superconductors, but also in the high-$T_c$ 
\cite{LondonHighTc} and heavy fermion species \cite{LondonHeavyFermion}.
It furnishes a generalization of the familiar phenomenon of Meissner 
screening to noninertial, material reference frames of the superconducting state.   
The London moment, then, represents a particularly striking instance of
a quantum protectorate \cite{laughlinpines}, for which the phenomenon
of macroscopic quantum coherence (the fact that the quantum of action 
$h=2\pi\hbar $ appears within a macroscopically measurable quantity)
``protects'' the bare property of a particle.
It is this phenomenon which enables the precise measurements of bare 
quantities related to the charge and mass of the electron. In the Quantum 
Hall effect, one measures in effect the fine structure constant $e^2/4\pi \epsilon_0\hbar c$, 
which can be determined to an accuracy of 0.1 ppb in comparative measurements between two 
Hall probes \cite{compareQH}. In superconducting quantum interference devices
(SQUID), the quantum of (Cooper pair) flux $\Phi_0 =h/2|e|$ is used
as a standard to measure magnetic fields with unprecedented precision. 
The properties of Josephson junctions also made possible, {\it e.g.}, the 
confirmation of constancy of the {\em electrogravitochemical} 
potential (as opposed to the conventional electrochemical potential 
without the inclusion of a gravitational contribution),   
in a circuit with two Josephson junctions separated 7.2 cm in height, which amounts to perpetuating 
a voltage constancy of $10^{-22}$ Volts over a time span of ten hours \cite{JainPRL87}.   
Correspondingly, due  to the fact that the ratio $m/2|e|$ is on a level of at least 
one ppm the bare ratio of the vacuum, a superconductor should be able to measure its own state of
rotation and, more general, acceleration with very high accuracy.
   
In the following, we give an account of the influence of noninertial forces
on macroscopic quantum devices, as specifically represented
by SQUIDs and Josephson junction arrays. This includes a study of  
the perfectly conducting and perfect diamagnetism analogues of the 
Tolman-Stewart respectively Barnett effects, the latter effect in the
superconductor being represented by the London moment.  
The prospects of detecting with high precision
the force fields related to noninertiality are given. 
  
The fact that the exact bare mass appears in the London equation, which
relates mechanical and magnetic quantities, implies that 
an effective theory, describing the motion of the massive current carriers,  
may be construed in a particularly transparent way. 
Specifically, in the linear in velocity, nonrelativistic limit and
for small deviations from the Minkowski metric of flat 
space-time, a general gauge invariance principle can be satisfied 
\cite{qhrotation}, which puts mechanical and proper electromagnetic
forces on an equal footing, uniting them into electromagnetomotive
forces. This program of generalized gauge invariance is described  
in the following section \ref{force fields} on the basis of nonrelativistic kinematics. 
In an appendix, we outline the derivation of this gauge invariance program
extracted from relativistic geodesic motion, and relate the potentials of the noninertial 
force fields to metric coefficients in weakly perturbed Minkowski space-time.
In section \ref{secJJA} we then describe the influence of electromagnetomotive force fields on 
the electrodynamics of superconducting quantum interferometers. In particular, the influence
of acceleration and rotation on the voltage response function of one-dimensional Josephson junction arrays
is discussed. For special Josephson junction arrays, so-called superconducting quantum interference filters
\cite{OHSSQIF,IEEE_SQIF,Appl_Serif},
it is explicitly shown how such devices can be used for precision measurements of rotation.
The knowledge of the electromagnetomotive force fields in the superconductor 
enables as an application the 
sensitive tracking of the {\em trajectory} of the quantum interference 
device. The procedure to be used for that purpose will be outlined in section \ref{Tracking}. 

\section{Electromagnetomotive force fields for accelerated samples in rigid body rotation}
\label{force fields}
\begin{center}
\begin{figure}[t]
\psfrag{Omega}{\large ${\bm \Omega}(t)$}
\psfrag{r}{\large ${\bm r}$}
\psfrag{v}{\large ${\bm v}$}
\psfrag{e}{\Large $q$}
\psfrag{r0}{\large ${\bm r}_0(t)$}
\psfrag{O}{$\rm O$}
\psfrag{JJ}{\large JJ}
\psfrag{T1}{\LARGE $\varphi_{\mbox{}_{L}}$}
\psfrag{T2}{\LARGE $\varphi_{\mbox{}_{R}}$}
{\includegraphics[width=0.4\textwidth]{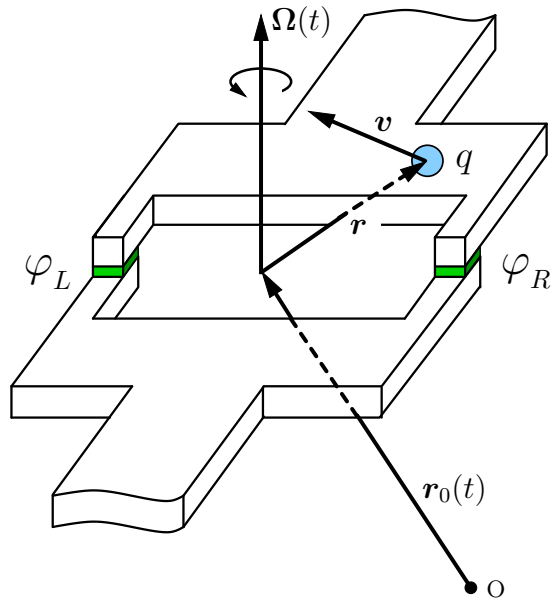}}
\vspace*{1em}
\caption{\label{squid} A Cooper pair in a rotating and accelerating  
superconducting quantum interference device (SQUID), with two Josephson junctions. 
Its position and velocity are given by their values $\bm r$ and $\bm v$ in the frame
rotating with angular velocity $\bm \Omega$ about a prescribed axis, which 
is located at a (time dependent) laboratory frame position ${\bm r}_0$. For representation 
purposes, the axis is in this picture located at the center of the SQUID and
perpendicular to its surface. $\varphi_{L}$ and $\varphi_{R}$ are the
gauge invariant phase differences associated to the Josephson junctions.}
\end{figure}
\end{center}

The noninertial force on a massive test particle inside a rotating and accelerating probe, 
as measured in the probe's rest frame, is given by the standard expression
\begin{eqnarray}
{\bm F}_{\rm noninertial} & = & 2m {\bm v}\times{\bm \Omega}-m {\bm \Omega}\times {\bm \Omega}\times {\bm r} 
- m \left( \partial_t{\bm \Omega} \right) \times {\bm r} \nonumber\\
& & - m\nabla \Phi -m \partial_t^2 {\bm r}_0
\label{noninertialforces} 
\end{eqnarray}
where $\bm \Omega$ is the rotation velocity; $-m\nabla\Phi $ is a possible scalar force on the particle, 
e.g., gravity, and $\Phi$ its potential. The first term on the right hand side of (\ref{noninertialforces}) 
represents (minus) the Coriolis force, the second one the centripetal force, and the third term is due to 
temporal changes of the angular velocity. The vector ${\bm r}_0$ is the position of the center of rotation, 
and $\partial_t^2 {\bm r}_0$ is an (externally imposed) linear acceleration of
this center of rotation (cf. Fig. \ref{squid}). 

Compare the relation (\ref{noninertialforces}) to the expression for the Lorentz force:
\begin{equation}
{\bm F}_{\rm Lorentz}= q {\bm v}\times {\bm B} + q {\bm E}\,, \qquad ({\rm Lorentz}) 
\end{equation} 
where as usual, provided that the conventional homogeneous Maxwell equations 
\begin{equation}
{\rm rot}\, {\bm E} = -\partial_t {\bm B}\,,\qquad {\rm div}\, {\bm B} =0   \label{convMaxwell} 
\end{equation}
hold, the magnetic and electric fields are derivable from vector and scalar 
potentials as
\begin{eqnarray}
{\bm E} =  \nabla A_0 -\partial_t {\bm A} \,,\qquad {\bm B} = \nabla \times {\bm A}\,.
\end{eqnarray}
We, then, define vector and scalar potentials associated

to noninertiality  as follows 
\begin{equation}
{\bm a}= {\bm \Omega} \times {\bm r}  + \partial_t {\bm r}_0 \,,\qquad a_0 = \frac12 \Omega^2 {\bm r}^2_\perp - \Phi\,,  
\label{aa0def}
\end{equation}
where ${\bm r}_\perp$ is the distance vector perpendicular to the axis of rotation.   
Summing the mechanical and electromagnetical forces, we may infer that for a charged massive particle 
like electron ($q=-\left|e\right|$) or Cooper pair ($q=-2\left|e\right|$), we can merge the above potentials and 
the electromagnetic potentials into a generalized vector potential, incorporating the coupling 
constants charge $q$ and mass $m$ \cite{mashhoon},   
\begin{equation}\label{Adef} 
{\cal A} = q{\bm A} + m {\bm a} \,,
\end{equation}
and a generalized scalar potential
\begin{equation}\label{chidef} 
{\chi} = -qA_0 - m a_0 \,.
\end{equation}
The sum of the generalized {\em electromotive} and {\em magnetomotive} forces, acting 
on an electron, consisting of noninertial plus proper Lorentz and electric forces, then takes on the form 
\begin{eqnarray}
{\bm F}_{{\cal L}} = {\cal E} +{\bm v}\times {\cal B}\,,
\label{genLorentz}
\end{eqnarray}
where the generalized electric and magnetic fields are given by the potentials $\cal A$ and $\chi$: 
\begin{eqnarray}
{\cal E} &=& -\nabla\chi -\partial_t {\cal A}
\nonumber\\
& = &q{\bm E} + m\nabla a_0 -m \partial_t {\bm a} 
\nonumber\\
& = & q{\bm E} -m {\bm \Omega}\times \left({\bm \Omega}\times {\bm r}\right) - m \left( \partial_t{\bm \Omega} \right)\times {\bm r} 
- m\nabla \Phi -m \partial_t^2 {\bm r}_0 \,,\nonumber\\
{\cal B} &=& \nabla \times {\cal A}\nonumber\\
& = & q {\bm B} + 2m{\bm \Omega} \,.\label{genFields}
\end{eqnarray}
As a consequence of relation (\ref{genLorentz}) for the total force, 
the usual expression for the drift velocity of the charge carriers, resulting from zero total force  
in perpendicular electric and magnetic fields, experiences the obvious modification that ${\bm E}\rightarrow {\cal E}$ 
and ${\bm B}\rightarrow {\cal B}$, so that ${\bm v}_D= {\cal E}\times {\cal B}/{\cal B}^2$.   

The generalized electromagnetic force fields displayed in equations (\ref{genLorentz}) and
(\ref{genFields}) give a theory possessing in effect two U(1) gauge symmetries. 
The standard U(1) from electromagnetism, with coupling constant $q$ (charge), 
and another U(1) gauge symmetry, with coupling constant $m$ (inertial rest mass).    
The gauge potential of this second U(1) has a scalar part $a_0$ and a vectorial part $a_i$. 
The homogeneous Maxwell equations 
\begin{eqnarray}
{\rm rot}\, {\cal E} & = & -\partial_t {\cal B},\label{Faraday}
\\
{\rm div}\, {\cal B} &= &0\label{noMonopole}
\end{eqnarray}
then follow from the existence of the generalized potentials 
$\cal A$ and $\chi$ in (\ref{Adef}) and (\ref{chidef}), which 
give the fields $\cal E$ and $\cal B$ in (\ref{genFields}).
They are identical to the conventional homogeneous 
Maxwell equations in (\ref{convMaxwell}) with the replacements
${\bm E}\rightarrow {\cal E}$ and ${\bm B}\rightarrow {\cal B}$.  
That the Faraday law (\ref{Faraday}) 
holds is due to our admitting a variation of the angular velocity with 
time and the resulting force term in (\ref{noninertialforces}). 
We stress that the fields $\cal E$ and $\cal B$ are both referring to the
frame co-rotating as well as co-moving with the quantum interference device
with respect to the laboratory frame. 
The laboratory frame velocity $\partial_t {\bm r}_0$ as well as
the rotation rate $\bm \Omega$ can be time dependent in an arbitrary manner. 

The gauge invariant particle (mass) current induced by the electromotive force 
field is in linear response 
\begin{equation}\label{sigmadef}
{\bm J}_i^{\rm ind} = {\bm {\tilde \sigma}}_{ij} {\cal E }_j \, .
\end{equation}
Observe that the left hand side contains the induced mass current density rather than the electric current density.
The associated response coefficient ${\bm {\tilde \sigma_{ij}}}$  is measured in units of $[\tilde \sigma ] = [\sigma_{el}/q^2]$. 
In the case of two coupling constants, $m$ and $q$, it is the 
number of particles crossing a unit area per unit time, which is 
the relevant observable. This quantity is proportional to the electromotive force field ${\cal E}$,
which causes these particles to move. 

Evidence for the necessity of using a {\em particle transport} equation in the form of (\ref{sigmadef}) 
comes from the existence of the London field in superconductors.  
Complete expulsion of the field $\cal B$ deep inside in a superconductor 
requires the particle conductivity $\bm {\tilde \sigma}$ to have a contribution proportional to 
$1/i\omega$, which yields a term on the right hand side
of (\ref{sigmadef}), proportional to the generalized vector potential 
$\cal A$. Corresponding to complete Meissner type screening,  
${\cal B}={\rm rot}\, {\cal A}= q {\bm B} + 2m{\bm \Omega}={0}$, 
the London spontaneous field ${\bm B}_L$ then takes the value 
\begin{equation}\label{London}
{\bm B}_L = -2\frac{m} q {\bm \Omega}\,.
\end{equation}
This relation corresponds to zero winding number of the 
phase $\theta$, cf. equations (\ref{pdqInt})--(\ref{PhiEq}) below. Equation  (\ref{London})  
was derived by F. London \cite{London1}, and
has been verified experimentally already 35 years ago \cite{zimmerman}, in an experiment  
in which it was used to infer the Compton wavelength of superconducting electrons. 
If we insert on the left hand side of the equation (\ref{London}) the bare electron values
$m=2m_e$ and $q = -2\left|e\right|$, we have 
\begin{equation}\label{BL}
\left|{\bm B}_L\right| = 7.15 \cdot 10^{-11}\,{\rm Tesla}
\end{equation}
for $\left|{\bm \Omega}\right|=2\pi{\rm /sec}$.
Quantum coherence properties are expressed by the requirement for the line integral 
of collective particle momentum along a closed path to be quantized: 
\begin{equation}\label{pdqInt}
\oint_{\cal C} \left\langle {\bm p}, d{\bm r}\right\rangle 
= N_v \, h \,,  
\end{equation}
where $N_v$ is the winding number of phase $\theta$,
so that the total canonical momentum 
\begin{eqnarray}
{\bm p} & \equiv &\hbar \nabla \theta \nonumber\\
& = & m{\bm v}_s  + {\cal A}\nonumber\\
& = & m{\bm v}_s + m \partial_t {\bm r}_0 + m {\bm \Omega}\times {\bm r}+ q {\bm A} \,,
\end{eqnarray}
where ${\bm v}_s$ is the Cooper pair velocity field.
It has a mechanical contribution proportional to $m$ and a proper electromagnetic
contribution $q{\bm A}$. The uniqueness condition of the collective
phase represented in (\ref{pdqInt}) then leads to the quantization 
of the sum of a Sagnac flux \cite{PostSagnac,stedman} and the magnetic flux
\begin{eqnarray} 
\Phi & = & \oint_{\cal C} \left\langle{\cal A}, d{\bm r}\right\rangle \\
& = & q\oint_{\cal C} \left\langle{\bm A}, d{\bm r}\right\rangle 
+ m \oint_{\cal C} \left\langle{\bm \Omega}\times {\bm r},d{\bm r}\right\rangle \nonumber\\
& = &  q\int \left\langle{\bm B},  d{\bm S}\right\rangle 
+ 2m \int \left\langle{\bm \Omega},  d{\bm S}\right\rangle\nonumber\\
& = & \int \left\langle{\cal B},d{\bm S}\right\rangle
= N_v \, h \,,  \label{PhiEq} 
\end{eqnarray} 
if we take a path in the bulk of the electron liquid, for which the integral of  
$m{\bm v}_s$ may be neglected.
This flux quantization rule associated with the field $\cal B$  
corresponds to the fact that a {\em vortex},
represented by a zero in the (collective) electron wave function,  
where the phase $\theta$ becomes singular, is  
fundamentally characterized by its winding number $N_v$ alone.  
No properties of the medium in which it lives, in particular the mass 
and charge of the medium's constituents, enter the quantum of generalized 
flux, which is given by Planck's quantum of action alone. 
The relation for the London moment in (\ref{London}), expressing vanishing magnetomotive force field, 
corresponds to zero winding number of the phase $\theta$. The classical 
property of zero generalized magnetic field ${\cal B} =0$
expressed by the London moment is hence rooted in the generalized Meissner
prescription $N_v=0$, and thus relates to the quantum coherence property expressed by (\ref{PhiEq}). 

The vanishing of the field ${\cal E}$ in the bulk of a noninertial superconductor (in the
static limit of zero frequency) implies that the proper electric field is nonzero inside the
superconductor, and given by
\begin{eqnarray}
{\bm E}_T & = & \frac{m}{q} \left[-\nabla a_0 +\partial_t {\bm a} \right] \label{EL}\\
& = & \frac{m}{q}\left({\bm \Omega}\times {\bm \Omega}\times {\bm r} 
+ \partial_t{\bm \Omega} \times {\bm r}+\partial_t^2 {\bm r}_0+ \nabla\Phi\right) \equiv -\frac{m}{q} {\bm g}\,.\nonumber
\end{eqnarray}
It is composed of the centrifugal, time variation of $\bm \Omega$, 
$\Phi$ potential and linear acceleration parts. If the total acceleration ${\bm g}$ is 
the gravitational acceleration on the surface of the earth, $\left|{\bf g}\right|=9.81 {\rm m}/{\rm sec}^2$, we have the value 
\begin{equation}
\left|{\bm E}_T\right| = 5.58 \cdot 10^{-11} {\rm Volt}/{\rm m}\,\label{ELexpl}
\end{equation}
for the electric field induced in the superconductor.
The total electromotive force field is thus simply 
\begin{eqnarray}
\label{electromotive}
{\cal E} & = & q ({\bm E} - {\bm E}_T)\nonumber\\
& = & q{\bm E} + m{\bm g} \,, 
\end{eqnarray} 
and takes a form analogous to the total magnetomotive force field 
\begin{eqnarray}
\label{magnetomotive}
{\cal B} & = & q ({\bm B} -{\bm B}_L)\nonumber\\
& = & q {\bm B} + 2m {\bm \Omega}\,.
\end{eqnarray}
The fact that there is an electric field associated to acceleration (which may be
material dependent for non-perfect conductors) was measured by Tolman and Stewart in metals \cite{tolmanstewart1}. 
The general phenomenon associated to $\cal B$, i.e., the occurrence of a magnetic field if the sample is rotated, 
was observed for magnetizable materials by Barnett \cite{barnett}, with a (possibly anisotropic) 
ratio of magnetic field and rotation different from the one displayed by superconductors expressible 
via the London equation (\ref{London}).  
The Tolman field ${\bm E}_T$ is a property of (in the limit of zero frequency) perfect conductors.
The London field ${\bm B}_L$, in turn, is a signature of perfect diamagnetism
for rotating samples.
Perfect diamagnetism may thus be understood as a  
hallmark of superconductors in general, be they considered in an 
inertial or noninertial reference frame. In short: The field ${\bm E}_T$ is measured if $\cal E$ is
vanishing and ${\bm B}_L$ is detected if $\cal B$ is completely (Meissner) screened.    

\section{Accelerated Josephson Junction Arrays} \label{secJJA}
\begin{center}
\begin{figure}[hbt]
\psfrag{Omega}{\large ${\bm \Omega}(t)$}
\psfrag{r}{\large ${\bm r}$}
\psfrag{r0}{\large ${\bm r}_0(t)$}
\psfrag{O}{O}
\psfrag{x}{\large x}
\psfrag{y}{\large y}
\psfrag{an}{\large ${\bf S}_{\mbox{}_{n}}$}
\psfrag{Tn}{\large $\varphi_{\mbox{}_{n}}$}
{\includegraphics[width=0.45\textwidth]{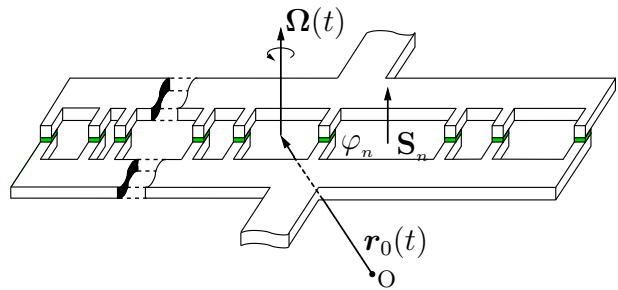}}
\vspace*{1em}
\caption{\label{sqif} Superconducting quantum interferometer with $N$ Josephson junctions. 
Its local frame rotates with angular velocity $\bm \Omega$ about a prescribed axis, which 
is located at a (time dependent) laboratory frame position ${\bm r}_0$. For representation 
purposes, the axis is in this picture located at the center of the interferometer and
perpendicular its plane surface. $\varphi_{n}$ is the
gauge invariant phase difference associated to the $n$th Josephson junction in the array and
${\bf S}_n$ denotes the $n$th orientated area element.}
\end{figure}
\end{center}

Devices based on superconductive quantum interference can be used as ultrasensitive detectors
for magnetic fields. They consist of one, two or even a plurality of Josephson junctions or weak links
which are connected as an array to form one or several superconducting loops. Prominent
exponents are devices containing two junctions per loop, like the dc-SQUID shown in Fig. \ref{squid}. It consists
of two junctions shunted parallel to form one single loop. Other devices of this class are series arrays of dc-SQUIDs 
or one-dimensional (1D) parallel arrays which can contain a plurality of Josephson junctions. A superconducting
interferometer based on an 1D parallel array is shown schematically in Fig. \ref{sqif}. 

\noindent
It consists of $N$ 
Josephson junctions shunted parallel in such a way that there are $N$-1
individual superconducting loops. In general the $N$-1 areas $S_n$ of these superconducting loops 
can have different shapes and sizes. In particular, judiciously choosing the distribution of the area loop sizes
in a suitable unconventional way, 1D parallel arrays can be used as sensors of absolute strength and 
orientation of magnetic fields. This is due to the fact that 1D parallel arrays are
magnetic field to voltage converters, if they are driven by a bias current of suitable strength.
Because of their unique response to applied magnetic fields, such 1D parallel arrays with unconventional grating
are named quantum interference filters (SQIF) and are explained in greater detail in \cite{OHSSQIF,IEEE_SQIF}. 

In the following the Josephson junctions are assumed to be {\it short} junctions such that any spatial variations
of the current density along the barriers of the weak links can be safely neglected. In this case each junction 
can be described by a gauge invariant phase difference
\begin{equation}
\label{invPhase}
\varphi=\theta_1-\theta_2+1/\hbar \int_1^2 \left\langle {\cal A}, d{\bf r} \right\rangle 
\end{equation}
of the macroscopic BCS pairing wave functions on either side of the weak link labeled 1 and 2 respectively.
Within the range of validity of the resistively and capacitively shunted junction (RCSJ) model \cite{Likharev}
the current through the Josephson junction $I$ is a superposition of the dissipationless macroscopic
supercurrent $I_{s}$ with a normal current, characterized by a shunt resistance $R$ and shunt capacitance $C$
\begin{equation}
\label{RCSJ}
I(\varphi)=\frac{\hbar C}{2\left|e\right|} \, \partial_t^2 \varphi+\frac{\hbar}{2\left|e\right|
R} \, \partial_t \varphi+I_{c}\sin(\varphi).
\end{equation}
For an ideal $S$-$I$-$S$ junction the supercurrent is connected to the phase difference $\varphi$ across the tunneling
barrier by $I_{s}=I_{c}\, \sin(\varphi)$, where $I_c$ is the maximum dissipationless supercurrent, that can 
flow through the junction. Of course, in 1D arrays each junction can have individual parameters $R_n$, $C_n$ and $I_{c,n}$.

In (\ref{invPhase}) there appears the generalized vector potential $\cal A$ from (\ref{Adef}) in the definition
of the gauge invariant phase difference. According to (\ref{genFields}), this indicates that, 
in principle, the electro- and magnetomotive force fields ${\cal E}$ and ${\cal B}$ can be measured by superconducting
quantum interferometers. For 1D parallel Josephson junction arrays, the basic relations are now discussed.

According to the fundamental Josephson relation the rate of change of the time dependent
phase difference $\varphi(t)$ is related to the electromotive force field across the junction barrier by
\begin{equation}
\label{josrel}
\hbar \partial_t \varphi(t)=-\int_1^2 \left\langle {\cal E}, d{\bf r} \right\rangle\,.
\end{equation}
In the case of a electromotive force field $\cal E$
the Josephson frequency $\nu$ for a single junction evaluates from the right 
hand side integral of (\ref{josrel}), which is the work functional 
associated with the electromotive force field:  
\begin{equation}
\label{freq}
h\nu = \hbar \, \lim_{t\rightarrow \infty} \frac{1}{t} \left[ \varphi(t) - \varphi(0) \right] \,.
\end{equation}
Scaling this in experimentally relevant units,  we have  that the
electric field (\ref{ELexpl}) induced by an acceleration $\left|{\bf g}\right|=9.81 {\rm m}/{\rm sec}^2$ corresponds to 
\begin{equation}
\nu_T = 27.0\, {\rm Hz} \;\; \frac{l}{\rm mm}, 
\end{equation}
where $l$ is the total length of the superconducting region 
(the length of the integration path in (\ref{freq}) joining
the two sides of the junction), in which the field ${\bm E}_T$ exists. 

Consider now the $n$th loop of the 1D array containing the junctions labeled $n$ and $n$+1, respectively. 
From (\ref{josrel}) it follows 
\begin{equation}
\hbar \frac{d}{dt} (\varphi_{n} -\varphi_{n+1})= -\oint_{{\cal C}_n} \left\langle{\cal E},d{\bm r}\right\rangle,\label{JosephsoncalV}
\end{equation}
where the path ${\cal C}_n$ circulates around the boundary of the surface element ${\bf S}_n$ just once.
The electromotive force field ${\cal E}$ determines via (\ref{JosephsoncalV}) the temporal evolution of the difference
of the variables $\varphi_{n}$ and $\varphi_{n+1}$ associated to the Josephson junctions in the considered loop.
These $N$-1 equations describe the effects of the electromotive force field ${\cal E}$ (cf. (\ref{genFields})) on 
accelerated 1D parallel arrays of Josephson junctions.

The basic formula describing the effects of magnetomotive effects on superconducting interferometers is the
condition of flux quantization. The generalized magnetic flux $\Phi_n$ through the area of the $n$th elementary 
loop ${\bf S}_n$ in an 1D parallel array determines via 
\begin{equation} 
\hbar( \varphi_{n}-\varphi_{n+1}) = \Phi_n=\int_{S_n} \langle {\cal B},d{\bf S} \rangle ,\label{ThetaFlux}
\end{equation}
the difference of the phase differences of the two junctions which form this loop. 
Taken severely this relation holds provided the superconducting loop is made of a material thick compared
to the magnetic penetration depth $\lambda$. In this case there exists a path inside the wire connecting
the junctions $n$ and $n$+1, on which the superfluid velocity field ${\bm v}_s$ becomes negligibly small.
So, $\hbar {\bf \nabla \theta}={\cal A}$ along this path.
In (\ref{ThetaFlux}) $\Phi_n$ is the generalized flux from (\ref{PhiEq}), incorporating both
the conventional magnetic flux and (twice) the flux of the rotation field.  
Therefore superconducting interferometers can in principle be used to determine the rate
of rotation $\Omega$ via the detection of the London spontaneous field ${\bf B}_L$
corresponding to $\Omega$ (cf. (\ref{London})).

By the generalized Faraday's law (\ref{Faraday}) the electromotive 
force field ${\cal E}$ along the integration path ${\cal C}_n$ 
that circulates the $n$th closed loop in the array just once is directly connected to the time derivative
of the flux threading this area element
\begin{equation}
\frac{d}{dt} \int_{S_n} \left\langle{\cal B} ,d{\bm S} \right\rangle= -\oint_{{\cal C}_n} 
\left\langle {\cal E},d{\bm r}\right\rangle\,. 
\end{equation}
A comparison of the time derivative of (\ref{ThetaFlux}) with (\ref{JosephsoncalV}) indicates that
these basic relations describing the effects
of electromagnetomotive force fields on 1D parallel arrays are consistent
with the generalized Faraday's law.

Using the RCSJ model (\ref{RCSJ}) and Kirchhoff's rule, the total current $I_b$ flowing through the array is obtained
as the {\it phase sensitive} superposition of the individual junction currents $I_n(\varphi_n)$
\begin{equation}
\label{kirchhoff}
I_b=\sum_{n=1}^{N}I_n(\varphi_n).
\end{equation}
The gauge invariant phase differences $\varphi_n$ of adjacent Josephson junctions in the array are not
independent, but are connected to each other by the condition of flux quantization (\ref{ThetaFlux}).
Neglecting the Biot-Savart type inductive couplings \cite{OHSSQIF} among
the currents flowing in the array, it follows from (\ref{ThetaFlux}) that one can eliminate all phase variables
$\varphi_n(t)$ in favor of a single phase variable, say $\phi(t)=\varphi_1(t)$. 
In this case (\ref{kirchhoff}) can be used to map the problem of $N$ coupled 
Josephson junctions onto a virtual {\it single} Josephson junction model and
there results a scalar (RCSJ-like) differential equation determining the phase difference $\phi(t)$ \cite{OHSSQIF}. 

The decisive quantity determining the response of the
1D parallel Josephson junction array on {\it magnetomotive} force fields (\ref{magnetomotive}) is the complex structure factor 
${\cal S}_N({\cal B})$ \cite{OHSSQIF}, given by  
\begin{equation}
{\cal S}_N({\cal B}) = \frac 1N \sum_{n=1}^{N} \frac{I_{c,n}}{I_c}\,\exp\left[
\frac{i}{\hbar} \sum_{m=0}^{n-1} \left\langle{\cal B}, {\bm S}_m \right\rangle\right]\,,   
\end{equation}
where the critical currents of the individual junctions are $I_{c,n}$
(their average over all $N$ junctions is $I_c$) and ${\bf S}_m$ are the orientated area elements
of the array (with ${\bf S}_0 = 0$).
The quantity ${\cal S}_N$ is strongly
affected by the geometry of the array, i.e. the choice of the individual area elements ${\bf S}_m$ 
(cf. Fig. \ref{sqif}), and describes interference effects between the array junction currents
in the presence of magnetomotive force fields.

In the overdamped junction regime and for static magnetomotive fields the scalar differential equation
of the single (virtual) Josephson junction model can be solved analytically 
under conditions where a constant current $I_b$ is biased. The solution $\phi(t)$ then determines 
via $V(t)=\hbar/2\left|e\right| \; \partial_t \phi(t)$ the voltage drop between the electrodes of the array.
It turns out \cite{OHSSQIF}, that if the bias current $I_b$ exceeds the maximal critical array current 
$N\,I_c \, |{\cal S}_N({\cal B})|$, the absolute value $|{\cal S}_N({\cal B})|$ influences the
time averaged voltage $\langle V \rangle$ across the array by
\begin{equation}
\frac{h\nu}{2\left|e\right|} = \langle V \rangle =\, I_c R \,\sqrt{\left(\frac{I_b}{N I_c}\right)^2-\left|{\cal S}_N ({\cal B})\right|^2} \, .
\label{voltresp}
\end{equation}
Here R denotes the average ohmic resistance of all array junctions.
Taking into account all inductive couplings, the qualitative behavior of the array voltage response
does get not affected, i.e., (\ref{voltresp}) also qualitatively describes the voltage response
in this case \cite{OHSSQIF}.

If the bias current $I_b$ is adjusted slightly above the array critical current,
the presence of magnetomotive force fields gives an effect of shifting the frequency (respectively the voltage)
which is orders of magnitude larger than the frequency shift displayed in (\ref{JosephsoncalV}).
The relevant quantity here is the maximum voltage transfer factor of the voltage response function:
\begin{equation}
{\cal T}_N=
\left|\frac{\partial \, \left(2\left|e\right| \, \left\langle\, V\,\right\rangle\right)}{\partial {\cal B}}\right|_{\rm max}\,
\end{equation} 
which determines the maximum sensitivity of the array on magnetomotive fields.
Scaling (\ref{voltresp}) in experimentally relevant units, we have for $\left|{\bm \Omega}\right|=2 \pi /{\rm sec}$ 
that the magnetomotive field (\ref{BL}), i.e. the London 
spontaneous field $\bm{B}_L$, corresponds to
\begin{equation}
\label{nuB}
\nu_L=3.46 \,\cdot 10^{4} \, {\rm Hz} \,\, \frac{{\cal T}_N}{\rm Volt/Tesla}\,
\end{equation}
provided the array is driven at its most sensitive point of operation. Typical experimental values for the transfer factor
of bare 1D parallel arrays (with $N=30$) are of the order of ${\cal T}_N\approx10^2$-$10^3$ Volt/Tesla \cite{IEEE_SQIF}. 
As can be derived from (\ref{voltresp}) the transfer factor scales with the number $N$ of junctions in the array, so that ${\cal T}_N$
can be increased with $N$. Using additional flux-focussing structures, e.g. superconducting pick-up loops, 
the transfer factor can be further increased by several orders
of magnitudes up to ${\cal T}_N\approx10^6$ Volt/Tesla. According to (\ref{nuB}), such devices are then very sensitive to
rotations and can measure the angular velocity $\bm \Omega$ very precisely.

{\centering
\begin{figure}[t]
\psfrag{V}{\huge $\langle V \rangle / (I_c R)$}
\psfrag{B}{\huge $\langle {\bf B}_L, {\bf S}_{\rm max}\rangle/\Phi_0$}
\resizebox*{0.95\columnwidth}{!}{\includegraphics{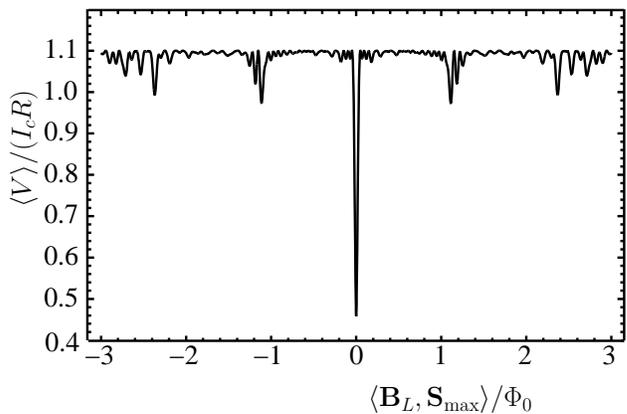}}
\vspace*{0.5em}
\caption{\label{figvoltresp} Voltage response due to rotation of a quantum interference filter, with $N=30$ (overdamped)
Josephson junctions for bias current $I_b=1.1\,N\,I_c$. The time averaged voltage $\langle V \rangle$ 
in units of $I_c\,R$ is plotted versus the normalized magnetic flux 
$\langle {\bf B}_L, {\bf S}_{\rm max}\rangle/\Phi_0$ which the London spontaneous field ${\bf B}_L$ induces in 
the largest area element ${\bf S}_{\rm max}$ of the array. The loop areas $\left|{\bf S}_n\right|$ are all in plane and randomly distributed 
between $0.1$ and $1.0 \,\left|{\bf S}_{\rm max}\right|$.}
\end{figure}}
In Fig. \ref{figvoltresp} the voltage response, according to (\ref{voltresp}), of a quantum 
interference filter due to rotation with angular velocity ${\bm \Omega}$ is shown. For vanishing magnetic field ${\bf B}=0$, the normalized voltage $\langle V \rangle/(I_c\,R)$ is plotted
versus the normalized magnetic flux $\langle {\bf B}_L, {\bf S}_{\rm max}\rangle/\Phi_0$ 
the London spontaneous field ${\bf B}_L=-2m/q \,{\bm \Omega}$ induces in the largest area element ${\bf S}_{\rm max}$ of the array.
The SQIF contains $N=30$ junctions and the loop areas $S_n=\left|{\bf S}_n\right|$ are randomly distributed 
between $0.1$ and $1.0 \,\left|{\bf S}_{\rm max}\right|$. For maximal voltage swing
the bias current $I_b$ is adjusted slightly above $N\,I_c$.

For vanishing magnetic field ${\bf B}=0$, the voltage response is indeed a {\it unique} function 
of the London field $\left|{\bf B}_L\right|$ and hence of $\left|{\bf \Omega}\right|$ around
its {\it global} minimum at $\left|{\bf B}_L\right|=0$. This suggests that it would be possible, e.g. by
measuring control current(s) flowing through the wires of suitably orientated compensation coil(s), to
reconstruct the absolute value, the orientation and even the phase of the rate of rotation, 
i.e. to determine the full angular velocity vector ${\bm \Omega}(t)$ and its time dependence.

A basic problem for the sensitive detection of rotation and other noninertial fields is that the
device has to be shielded against external magnetic fields, like
that of the earth. It has been shown in Ref. \cite{gyroshield} 
by Satterthwaite and Gawlinski for the
stationary case, that superconducting shielding, which delivers the shielding
factors required to detect rotations as slow as, e.g., that of the earth,
implies that the apparatus cannot distinguish
between applied rotation and applied proper magnetic flux: 
The co-rotating superconducting shield prevents such a distinction. 
The current induced by a rotation with $\bm \Omega$ is the same as that induced by an external, applied 
magnetic field $-{\bm B}_L$, because the device cannot tell from 
which of the two parts of $\cal B$ the induced current actually comes from.   
Shielding is not necessary (and, indeed, not possible) for charge 
neutral quantum interference devices \cite{AVearth}, which are thus
capable of detecting absolute rotation, whereas superconductively
shielded SQUIDs or Josephson arrays are not. 
What {\em is} measurable by superconducting interferometers are 
the fields $\cal B$ and $\cal E$ coming from the (accelerated) motion relative 
to the external shield, which remains fixed with respect to the local frame of inertia.
In particular, SQIFs can measure relative motion on an absolute scale.
However, a gyromagnetic gyroscope based on this idea needs some mechanics and therefore
can not be more sensitive than a mechanical gyroscope itself.

One possibility to build a gyromagnetic gyroscope without any moving parts, is to
use a material with a magnetomechanic factor $\gamma$ different from the factor $\gamma = -2m/q$, 
occurring in the London equation (\ref{London}) for the superconducting shield. In this case it is possible to 
circumvent the problem of indistinguishability, i.e. 
to which parts of $\cal B$ an induced current is related.
For example a ferromagnetic material, whose magnetic field induced by rotation
is itself measured by a SQUID which is shielded by a 
superconducting shield, can measure the rotation field \cite{vitale89}.
But such a device is not a superconducting gyroscope in a narrower sense  \cite{gyroshield}
which exclusively relies on the superconductors' response to rotation.

\section{Trajectory tracking}\label{Tracking}
The determination of the trajectory $x^i(t)+x^i_0(t)$ of an electron (or 
Cooper pair) in an accelerated superconductor amounts to solving 
the second order differential equation 
\begin{equation}
\frac{d^2  x^i}{dt^2} +\frac qm \epsilon_{ijk} ({\bm B}_L)_k \,\frac{d x^j}{dt} \label{ELBLGeodEq}
+\frac qm ({\bm E}_T)_i = 0 \,,
\end{equation} 
which is (\ref{noninertialforces}) with $\bm F= m\, d^2 x^i/dt^2$, 
where the in general time dependent proper magnetic and electric fields 
${\bm B}_L$ and ${\bm E}_T$ are determined from Eqs. (\ref{London}) and (\ref{EL}). 
The electric field ${\bm E}_T$ is, according to (\ref{EL}), in 
the rotational part position dependent, linear in the distance vector ${\bm r}=x^i {\bm e}_i$ 
from the center of (local) solid body rotation \cite{QMLinearR}.
In the appendix, equation (\ref{ELBLGeodEq}) is explained in terms the 
geodesic equation (\ref{geodesic}), and the fields ${\bm E}_T$ and ${\bm B}_L$ are
identified as connection coefficients on a Riemannian manifold, {\it i.e.}, 
on a manifold representing space-time with some metric coefficients. 
 
\section{Conclusions}
Moving superconducting circuits consisting of current biased superconducting
quantum interference filters (SQIFs) are local oscillators that undergo a
characteristic and unique change of their oscillation frequency under
acceleration and rotation. However, by the very nature of the combined
vector field ${\cal A}$, Eq.(\ref{Adef}), such a superconducting interferometer
is only capable to detect the combined magnetomotive field ${\cal B}$,
Eq.(\ref{magnetomotive}), and the combined electromotive field ${\cal E}$,
Eq.(\ref{electromotive}). Employing a suitable shield such SQIFs
can nevertheless measure relative motion on an absolute
scale. The aforementioned devices might be used, for example, to construct
an absolute detector of noninertial motion in the context of seismology.

\appendix
\section*{Noninertial force fields from the geodesic equation}
The motion of a test particle, upon which no external (electromagnetic) 
force is acting, is describable by the geodesic equation in space-time, 
\begin{equation}\label{geodesic}
\frac {du^\mu}{d\tau} +\Gamma^\mu{}_{\alpha\beta} u^\alpha u^\beta=0\,,  
\end{equation}
with the four-velocity normalized to unity, {\it i.e.} $u^\mu u_\mu =-1$.
The connection coefficients $\Gamma^\mu{}_{\alpha\beta}$ serve to describe 
any kind of ``acceleration'' $d u^\alpha/d\tau$, caused 
by the transformation to the curvilinear co-ordinates of a rotating 
and accelerating frame (cf. equation (\ref{noninertialforces}) valid for a rigidly rotating frame), and 
nonrelativistic particle velocities. The true, invariant four-acceleration 
is invariantly zero for a geodesic: The equation above describes the 
(kinematic) autoparallel property of the four-velocity with components $u^\alpha$.
If electromagnetic fields are present, the right hand side of (\ref{geodesic})
is no longer zero, and  the covariant Lorentz force four-acceleration 
equation on a particle of charge $q$ and 
inertial rest mass $m$ in the presence of an electromagnetic field acting on the particle reads
\begin{equation}
\label{Lorentzacceleration}
m\frac{du ^\mu}{d\tau} 
=\left(q F^\mu{}_\nu - m\Gamma^\mu{}_{\alpha\nu} u^\alpha\right) u^\nu\,, 
\end{equation} 
where we have brought the connection coefficient term to the right hand side.
  
In the weak field limit $g_{\mu\nu}\simeq \eta_{\mu\nu} + h_{\mu\nu}$, in which raising and lowering of indices 
is to lowest order in $|h_{\mu\nu}|\ll |\eta_{\mu\nu}|$ accomplished by 
$\eta_{\mu\nu}= {\rm diag}\,(-1,1,1,1)$, the connection coefficients take on the form
\begin{eqnarray}
\Gamma^\mu{}_{\alpha\nu}=\frac12 \eta^{\mu\beta}\left( h_{\beta\alpha,\nu} + h_{\beta\nu,\alpha} - h_{\alpha\nu,\beta}\right) 
\,. \label{hconnection}
\end{eqnarray}
We now use that the spatial components of (\ref{geodesic}) are \cite{weinberggravity}
\begin{eqnarray}
\frac{d^2x^i}{dt^2} & = &-\Gamma^i{}_{00} -2 \Gamma^i{}_{0j} \frac{d x^j}{dt} -\Gamma^i{}_{jk}  \frac{d x^j}{dt}  \frac{d x^k}{dt}\nonumber\\
& & +\left[\Gamma^0{}_{00} +2 \Gamma^0{}_{0j}\frac{d x^j}{dt}+\Gamma^0{}_{jk}\frac{d x^j}{dt} \frac{d x^k}{dt} \right] \frac{d x^i}{dt}\nonumber\\
&\simeq & -\Gamma^i{}_{00} -2 \Gamma^i{}_{0j} \frac{d x^j}{dt} \,. 
\label{longgeodesic} 
\end{eqnarray}
The last line holds if we consider the lowest (linear) order in 
the charge velocity $\bm v$, whose magnitude is assumed to be much less than the speed of light. 
The rotation rate $\bm \Omega$, {\it i.e.}, the invariant (vorticity) measure of the proper velocity 
${\bm \Omega} \times {\bm r}$ induced by rotation, is taken into account up to $O({\bm \Omega}^2)$, 
 in the form of $\Gamma^i{}_{00}$. Terms which are quadratic in $\bm v$ (last term in the first line of (\ref{longgeodesic})),  
and those of higher order than quadratic in $\bm \Omega$ and $\bm v$
and their products (terms in the second line of (\ref{longgeodesic})) are neglected.  

Equation (\ref{Lorentzacceleration}) then gives the following spatial components 
\begin{eqnarray}
m \frac{d^2x^k}{dt^2} &= & q F_{k 0}-m\Gamma^k{}_{00} +(q F_{k i} -2m  \Gamma^k{}_{0i}) v^i\nonumber\\ 
&= & q (F_{k 0} +F_{k i}v^i)-m\left(   h_{k 0,0} - \frac12 {h_{00,k}}\right)  \nonumber\\
& & \qquad -m ( h_{k 0,i} + h_{k i,0} - h_{0i,k})v^i\,.
\label{linearForce}
\end{eqnarray}
This relation leads to the generalized {\em electromotive}
and {\em magnetomotive} force fields, reinstating the speed of light $c$, and neglecting the time derivative of $h_{ik}$, 
\begin{eqnarray}\label{genEEq}
{\cal E}& = & q{\bm E} +\frac12 mc^2 \nabla h_{00} - mc\partial_t{{\bm h}}_0\,,\\
\label{genBEq}
{\bm v}\times{\cal B} & = &{\bm v}\times \left( q{\bm B} + mc\nabla \times {\bm h}_0\right)\,.
\end{eqnarray}

The relations (\ref{genBEq}) result in the following identifications  with  the potentials in (\ref{aa0def}):  
\begin{equation}
\frac12 c^2 h_{00} = a_0 \,, \qquad c{\bm h}_0 = {\bm a}\,.
\end{equation}
Measuring the fields ${\bm B}_L$ and ${\bm E}_T$ in (\ref{London}) and 
(\ref{EL}) thus yields the connection coefficients $\Gamma^i{}_{00}$ and 
$\Gamma^i{}_{0j}$ (spatial coordinates are in a Cartesian frame):  
\begin{eqnarray}
({\bm E}_T)_k & = & \frac{m}{q} \Gamma^k{}_{00}\,,\nonumber\\
({\bm B}_L)_j \epsilon_{ikj} &=& -2\frac{m} q ({\bm \Omega})_j\epsilon_{ikj} =
 -\frac{2m}{q} \Gamma^k{}_{0i} 
\end{eqnarray}
For the distinction and understanding of ``real'' electromagnetism and generalized
electromagnetism as expounded here, it is of importance to bear in mind 
that ${\bm E}_T$ and ${\bm B}_L$, if understood as connection 
coefficients like in the relations above, have no {\em exact}
co-ordinate invariant meaning as tensor fields like the proper electromagnetic 
fields $\bm E$ and $\bm B$ have. 
They gain an {\em approximate} co-ordinate invariant meaning only in the 
weak field limit $|h_{\mu\nu}|\ll |\eta_{\mu\nu}|$, and if 
$\partial_t h_{ik}$ is negligible, because in this limit
the $\Gamma^{\alpha}{}_{\beta\gamma}$ in (\ref{hconnection}) transform tensorially, 
and the field strength is identified to be 
\begin{eqnarray}
-\Gamma_{\mu 0 \nu} & = & \Gamma_{\nu 0 \mu} =  
-\frac12\left( h_{0\mu,\nu} + h_{\mu\nu,0}- h_{0\nu,\mu}\right) \nonumber\\
&= &  {\cal F}_{\mu\nu} =   
\partial_\mu {\cal A}_\nu - \partial_\nu {\cal A}_\mu\,.
\end{eqnarray}    
In the specified limit of small velocities and small deviations from Minkowski space-time, we can thus ascribe 
co-ordinate invariant, {\it i.e.} tensorial  meaning to ${\cal F}_{\mu\nu}$. 
 

\end{document}